\documentclass[aps,12pt,preprint,notitlepage,oneside,onecolumn,nobibnotes,nofootinbib,superscriptaddress,noshowpacs,centertags]{revtex4}
\usepackage{epsfig}
\usepackage{graphics}
\begin{document}
\title{New Mass and Lifetime Measurements of $^{152}$Sm Projectile Fragments with Time-Resolved Schottky Mass Spectrometry}
\author{Yu.A.~Litvinov}%
\affiliation{Gesellschaft f\"ur Schwerionenforschung GSI,
Planckstra{\ss}e 1, 64291 Darmstadt, Germany} \affiliation{II.
Physikalisches Institut, JLU Giessen, Heinrich-Buff-Ring 16, 35392
Giessen, Germany}%
\author{F.~Bosch}%
\affiliation{Gesellschaft f\"ur Schwerionenforschung GSI,
Planckstra{\ss}e 1, 64291 Darmstadt, Germany}%
\author{H.~Geissel}
\affiliation{Gesellschaft f\"ur Schwerionenforschung GSI,
Planckstra{\ss}e 1, 64291 Darmstadt, Germany} \affiliation{II.
Physikalisches Institut, JLU Giessen, Heinrich-Buff-Ring 16, 35392
Giessen, Germany}%
\author{H.~Weick}%
\author{K.~Beckert}%
\author{P.~Beller}%
\affiliation{Gesellschaft f\"ur Schwerionenforschung GSI,
Planckstra{\ss}e 1, 64291 Darmstadt, Germany}%
\author{D.~Boutin}
\affiliation{Gesellschaft f\"ur Schwerionenforschung GSI,
Planckstra{\ss}e 1, 64291 Darmstadt, Germany} \affiliation{II.
Physikalisches Institut, JLU Giessen, Heinrich-Buff-Ring 16, 35392
Giessen, Germany}%
\author{C.~Brandau}
\affiliation{Gesellschaft f\"ur Schwerionenforschung GSI,
Planckstra{\ss}e 1, 64291 Darmstadt, Germany}\affiliation{Institut
f{\"u}r Atom- und Kernphysik, JLU Giessen, Leihgesterner Weg 217,
35392
Giessen, Germany}%
\author{L.~Chen}
\affiliation{II. Physikalisches Institut, JLU Giessen,
Heinrich-Buff-Ring 16, 35392 Giessen, Germany}%
\author{O.~Klepper}%
\affiliation{Gesellschaft f\"ur Schwerionenforschung GSI,
Planckstra{\ss}e 1, 64291 Darmstadt, Germany}
\author{R.~Kn{\"o}bel}
\affiliation{Gesellschaft f\"ur Schwerionenforschung GSI,
Planckstra{\ss}e 1, 64291 Darmstadt, Germany} \affiliation{II.
Physikalisches Institut, JLU Giessen, Heinrich-Buff-Ring 16, 35392
Giessen, Germany}%
\author{C.~Kozhuharov}%
\author{J.~Kurcewicz}%
\affiliation{Gesellschaft f\"ur Schwerionenforschung GSI,
Planckstra{\ss}e 1, 64291 Darmstadt, Germany}%
\author{S.A.~Litvinov}%
\affiliation{Gesellschaft f\"ur Schwerionenforschung GSI,
Planckstra{\ss}e 1, 64291 Darmstadt, Germany} \affiliation{II.
Physikalisches Institut, JLU Giessen, Heinrich-Buff-Ring 16, 35392
Giessen, Germany}%
\author{M.~Mazzocco}%
\affiliation{Gesellschaft f\"ur Schwerionenforschung GSI,
Planckstra{\ss}e 1, 64291 Darmstadt, Germany}%
\author{G.~M{\"u}nzenberg}
\affiliation{Gesellschaft f\"ur Schwerionenforschung GSI,
Planckstra{\ss}e 1, 64291 Darmstadt, Germany}
\affiliation{Institut f{\"u}r Physik, JGU Mainz, Staudingerweg 7,
55099 Mainz, Germany}%
\author{C.~Nociforo}%
\author{F.~Nolden}
\affiliation{Gesellschaft f\"ur Schwerionenforschung GSI,
Planckstra{\ss}e 1, 64291 Darmstadt, Germany}%
\author{W.~Pla{\ss}}
\affiliation{II. Physikalisches Institut, JLU Giessen,
Heinrich-Buff-Ring 16, 35392 Giessen, Germany}%
\author{C.~Scheidenberger}%
\author{M.~Steck}%
\author{B.~Sun}
\author{M.~Winkler}
\affiliation{Gesellschaft f\"ur Schwerionenforschung GSI,
Planckstra{\ss}e 1, 64291 Darmstadt, Germany}%
\begin{abstract}
The FRS-ESR facilities at GSI provide unique conditions for
precision measurements with stored exotic nuclei over a large
range in the chart of nuclides. In the present experiment the
exotic nuclei were produced via  fragmentation of $^{152}$Sm
projectiles in a thick beryllium target at 500-600 MeV/u,
separated in-flight with the fragment separator FRS, and injected
into the storage-cooler ring ESR. Mass and lifetime measurements
have been performed with bare and few-electron ions. The
experiment and first results will be presented in this
contribution.
\end{abstract}
\keywords{}
\maketitle
\section{Introduction}
\par%
The mass and the half-life of a nucleus are fundamental properties
which result from the interaction of all nucleons \cite{Bohr}.
Both quantities  are essential for the understanding of nuclear
structure and also for the nucleosynthesis in astrophysics.
\par
New phenomena in nuclear physics on shell structure, pairing
correlations, etc. have been discovered with precise nuclear
masses. The driplines, which determine the borders of nuclear
existence, are obtained from the mass differences of neighboring
nuclei. The actual paths of the nucleosynthesis in stars are
governed by the nuclear binding energies and lifetimes.
\par
A very important motivation for measuring new masses of exotic
nuclides is the test and improvement of nuclear theories. Although
the progress of the theoretical calculations was enormous in the
last years \cite{{Bender_RMP},{Lunney_RMP}}, especially in the
microscopic calculations, their predictive power is still up to a
factor of 100 worse compared to our presently achieved
experimental accuracy \cite{Li-2005, Li-PRL}.
\par
In this contribution we report on first results from a very recent
experiment performed at GSI. The experiment consisted of two
parts. In the first part the lifetimes of stored hydrogen-like
$^{140}$Pr$^{58+}$ ions have been measured and the objective of
the second part were the masses of neutron-deficient $^{152}$Sm
projectile fragments.

\section{Experiment}
\subsection{Production, separation, storage, and frequency measurement}
\par
\begin{figure}[t!]
\includegraphics*[width=\textwidth]{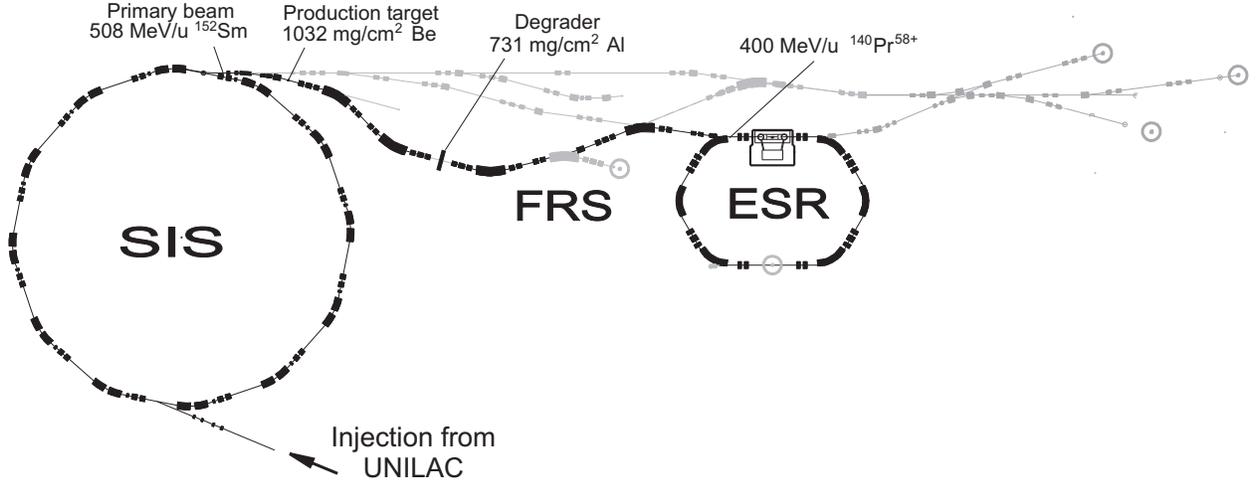}
\caption {Schematic layout of the secondary nuclear beam facility
at GSI. The heavy-ion synchrotron SIS, the fragment separator FRS,
and the storage-cooler ring ESR used in this experiment are
highlighted. The indicated primary beam energy, the production
target, degrader, and the energy of the injected into the ESR
fragment are those used in the first part of the experiment
devoted to the lifetime measurements of stored hydrogen-like
$^{140}$Pr$^{58+}$ ions.} \label{facility}
\end{figure}

%
Proton-rich nuclides were produced via fragmentation of the 508
and 615~MeV/u $^{152}$Sm primary beam, provided by the heavy-ion
synchrotron SIS \cite{Blasche}, in the 1032 and 4009~mg/cm$^2$
$^9$Be production targets, respectively. The first combination was
used for the lifetime measurements of $^{140}$Pr$^{58+}$ ions and
the second one was used in the mass measurements. The target was
placed in front of the fragments separator facility FRS
\cite{Ge-NIM24}. The fragments were separated in flight and
injected, stored, and electron cooled in the storage ring ESR
\cite{Franzke}. The experimental facility is schematically
presented in Figure~\ref{facility}. The experimental conditions
used in the first part of the experiment are indicated in the
figure.
\par%
The injection into the ESR was optimized with the primary beam and
the electro-magnetic fields of the FRS-ESR facilities were set at
a constant magnetic rigidity value during the measurements. The
magnetic rigidity value of $B\rho$=7.655~Tm was required in the
first part of the experiment and $B\rho$=6.5~Tm was used in the
part of the experiment aimed at mass measurements of
neutron-deficient nuclei. In general, the different selections of
the secondary fragments were done by changing the primary beam
energy impinging on the production target. The selected reference
fragment emerging from the target has then to match the prepared
ion-optical setting. In this experiment our reference fragment was
$^{108}$Sb$^{51+}$. In principle, all fragments in same magnetic
rigidity acceptance are transmitted as well. This is successfully
used in our mass measurements by simultaneously storing the
nuclides with unknown masses and with known masses for
calibration.
\par%
However, applying another separation criterion in addition we can
easily reduce and further select the number of nuclear species
injected into the ESR. Such a separation is possible with atomic
energy-loss in matter and a two-fold magnetic rigidity analysis,
B$\rho$-$\Delta$E-B$\rho$ method \cite{Ge-NIM24}. The
B$\rho$-$\Delta$E-B$\rho$ method was applied in this experiment
for half-life measurements of hydrogen-like $^{140}$Pr fragments.
In this case we want to avoid that neither the mother nor the
daughter nuclei are contaminated by other fragments, such as e.g.
helium-like $^{140}$Nd$^{58+}$ ions. Moreover, in order to obtain
the exact number of $^{140}$Pr$^{58+}$ ions decaying via nuclear
electron capture to $^{140}$Ce$^{58+}$ ions, the amount of
injected daughter ions should be kept as small as possible.
731~mg/cm$^2$ aluminium degrader was used at the middle focal
plane of the FRS (see Figure~\ref{facility}). The first half of
the FRS before the degrader was set to transmit fully-ionized
$^{140}$Pr$^{59+}$ ions. By applying this FRS setting, and also
using the slit systems, no $^{140}$Ce$^{58+}$ ions can be
transmitted till the degrader. The atomic charge state
distribution after the production target is very similar for
praseodymium and neodymium and amounts to about 86\% in the fully
ionized state, about 13\% in the hydrogen-like state, and about
0.5\% in the helium-like state (GLOBAL \cite{GLOBAL}
calculations). The corresponding charge state distributions after
the degrader are also very similar to the one above (the degrader
thickness is well above the equilibrium thickness of about
255~mg/cm$^2$ \cite{GLOBAL}). Thus, setting the second half of the
FRS and the ESR on the wanted $^{140}$Pr$^{58+}$, we achieved that
the intensity of $^{140}$Nd$^{58+}$ in the ESR was less than 1 per
mill of the praseodymium intensity. No other fragments have been
transmitted in this setting.
\par%
The ions injected and stored in the ESR were electron cooled. The
electron cooling process contracts the phase-space volume of
stored beams and the initial velocity distribution is reduced to
typically $\Delta{v}/v\approx5\times10^{-7}$. At the injection
only 25\% of the ESR acceptance is filled but after electron
cooling the circulating ions have exactly the same mean velocity
and thus occupy the entire storage acceptance of about $\pm$1.2\%
\cite{Raidi} due to their different mass-to-charge ratios. By
selecting the voltage of the cooler electrodes we define the
velocity of the merged electrons and thus the velocity of the
cooled ions.
\par
Besides the electron cooling the ESR is also equipped with a
stochastic cooling device \cite{No-NIMA} which provides fast
precooling at a fixed fragment velocity, corresponding to 400
MeV/u, and allows to access shorter-lived nuclei as demonstrated
in our previous experiments~\cite{Ge-RNB6}. This fixed velocity
results in a magnetic rigidity of $B\rho$=7.655~Tm in the part of
the run devoted to lifetime studies in which we have applied the
stochastic cooling. However, for the mass measurements we reduced
the magnetic rigidity of the ESR to stay in the optimum operating
domain of the electron cooler which was employed throughout the
present experiment.
\par%
The masses and lifetimes have been measured with the time-resolved
Schottky mass spectrometry (SMS) \cite{Li-2005}. It is based on
the Schottky-noise spectroscopy \cite{Borer}, which is widely used
for non-destructive beam diagnostics in circular accelerators and
storage rings. The stored ions were circulating in the ESR with
revolution frequencies of about 2~MHz. At each turn they induced
mirror charges on two electrostatic pick-up electrodes. The
30$^{th}$-31$^{st}$ harmonics of the signals were down-shifted to
the frequency range from 0 to 300~kHz, digitized with a 640 kHz
sampling rate, and stored as 16-bit words on a hard-disk for the
off-line analysis.
\par%
Fast Fourier Transformation is applied to the stored data leading
to the revolution frequency spectra. The frequencies provide
information about the mass-over-charge ratios of the
ions~\cite{{Raidi},{Li-2005}}. The area of the frequency peak is
proportional to the number of stored ions, which is the basis for
lifetime measurements \cite{Ge-RNB6}. The details of the data
acquisition system and of the data analysis can be found in
Ref.~\cite{Li-2005} and references therein.
\subsection{Magnetic rigidity acceptance of the ESR}
\par%
\begin{figure}[b!]
\includegraphics*[width=14cm]{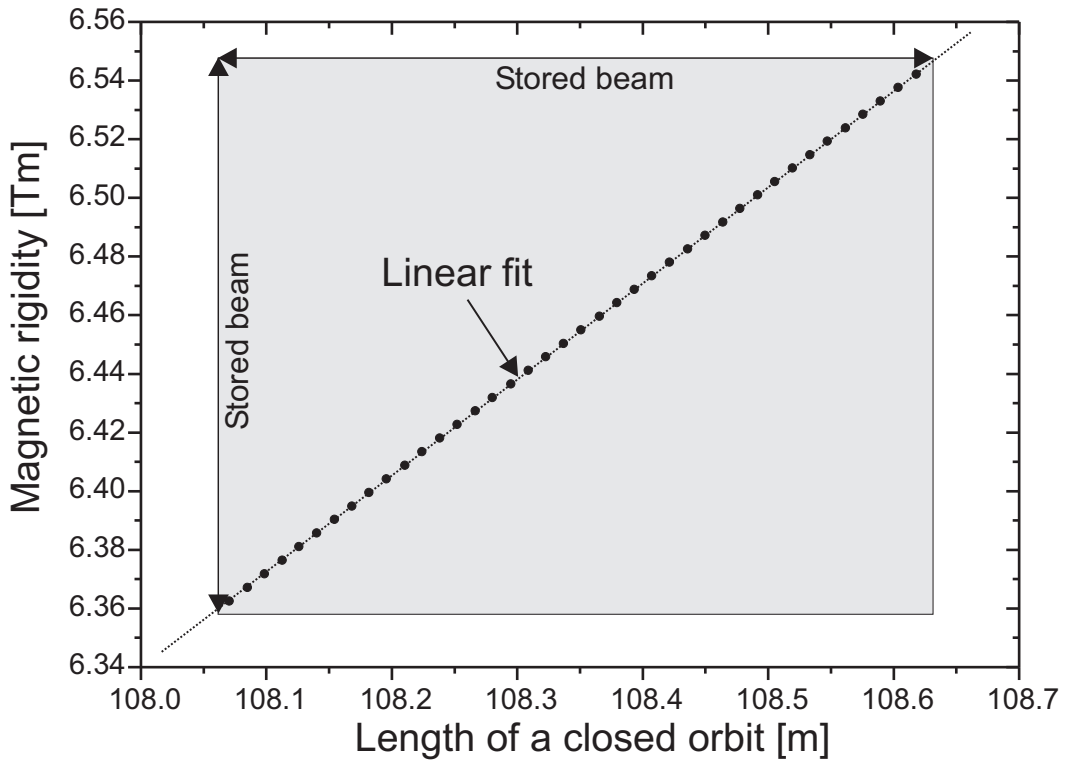}
\caption {Calibration of the ESR acceptance with the $^{152}$Sm
primary beam at different velocities (see text). Beyond the
presented data points no orbiting ions were observed.}
\label{calibration}
\end{figure}
It is important to know the range of the mass-over-charge values
which can be simultaneously stored in the ESR with given electron
cooler parameters such as electron current and cooler voltage.
Therefore, we have measured the $B\rho$ acceptance of the ESR.
\par%
The  $B\rho$ value of any stored ion is defined by its
mass-over-charge ratio $m/q$ and its velocity $v$ via:
$$
B\rho=\frac{mv\gamma}{q},
$$
where $\gamma$ is the relativistic Lorentz factor. For this
calibration measurement we used the primary beam
$^{152}$Sm$^{62+}$ ions with precisely known mass-over-charge
ratio \cite{AW03}. Since the velocity is defined by the electron
cooler voltage and electron current, the magnetic rigidity can be
determined.
\par%
The revolution frequency $f$ of the primary beam has been measured
with SMS for different beam velocities and the length of the
corresponding closed orbit $L$ was then calculated via $f=v/L$.
\par%
The experimental data points are shown in
Figure~\ref{calibration}. The error bars of each point are within
the symbol size. The linear fit was used to parameterize the data.
\par%
It is obvious that with this calibration one can exactly select
the measured mass-over-charge range of stored fragments by varying
the cooler parameters.
\section{Preliminary results}
\subsection{Mass measurements of neutron-deficient nuclides}
\par%
The present mass accuracy of the time-resolved SMS is about
30~$\mu{u}$ \cite{Li-2005}. Therefore, the objective of this
experiment was the mass surface which is presently unknown or
experimentally known but with error bars larger than the SMS
accuracy. The present status of knowledge of nuclear masses was
taken from the Atomic Mass Evaluation (AME) 2003~\cite{AW03}.
\par%
The production yields, ionic charge-state distribution,
transmission through the FRS, and injection into the ESR have been
calculated with the MOCADI~\cite{MOCADI} and the
LISE++~\cite{LISE} codes. The optimum setting was obtained for
$^{108}$Sb$^{51+}$ ions being at the magnetic rigidity of 6.5~Tm
corresponding to the central orbit of the FRS.
\par%
Using the calibration curve from Figure~\ref{calibration}, the
cooler voltages were calculated which are needed to cover the mass
surface aimed at in the experiment. The voltage was varied in
steps of 2~kV from 190~kV till 220~kV. The cooler current was kept
constant at 0.4~A, a relatively high current chosen for fast
electron cooling. With these parameters, the nuclei with
half-lives longer than one second are expected to be recorded in
the frequency spectra.
\par
\begin{figure}[h!]
\includegraphics*[width=\textwidth]{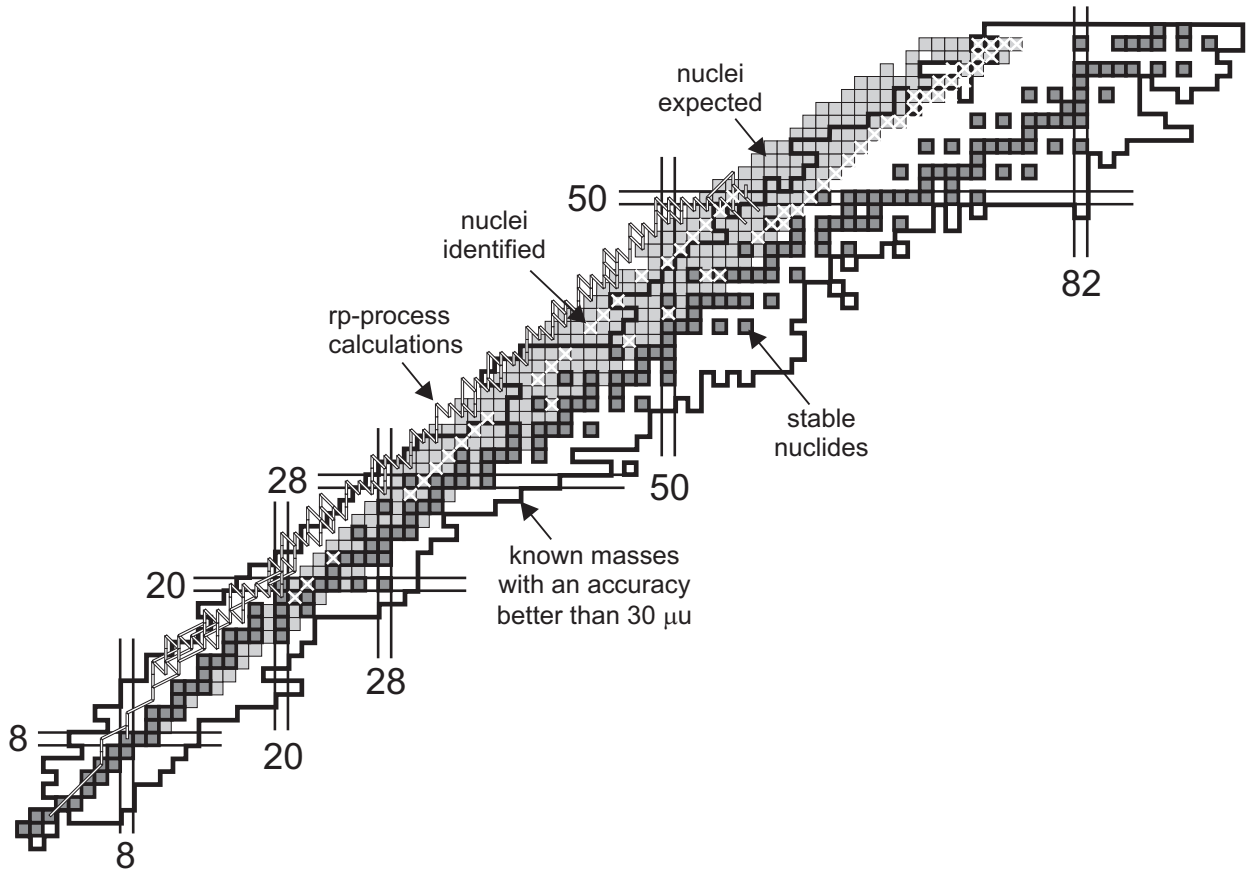}
\caption {Part of the chart of nuclides showing the mass surface
which is expected to be covered in this experiment. The
rp-path~\cite{rp}, the stable nuclei, and the nuclides with very
well known mass values~\cite{AW03} are indicated in the figure.
Nuclides so far identified in the frequency spectra of this work
are indicated with white crosses.} \label{nchart}
\end{figure}

%
A part of the chart of nuclides with the mass surface expected to
be covered in this experiment is shown in Figure~\ref{nchart}. The
developed single particle method~\cite{Li-2005,GL-2005} is the
base for precise mass determination of even a single stored ion.
Thus, particle yields in the ESR as low as one ion in one hundred
injections could be measured. The analysis of the data is in
progress. The presently identified nuclides are indicated in the
figure with white crosses.
\par%
One can see from Figure~\ref{nchart} that the expected mass
surface as well as some of the presently identified nuclides lie
close to the calculated rp-process path~\cite{rp}.
\par
\begin{figure}[h!]
\includegraphics*[width=\textwidth]{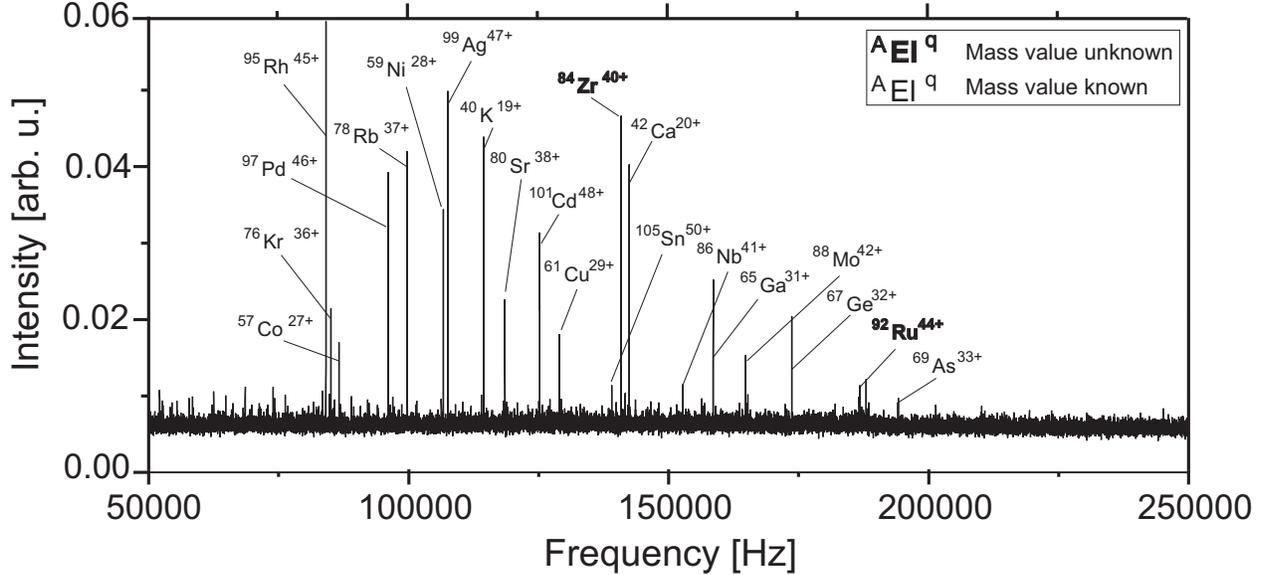}
\caption {Part of a Schottky frequency spectrum. Known and unknown
masses are indicated according to the AME 2003~\cite{AW03}.}
\label{spectrum}
\end{figure}

%
An example of a measured Schottky frequency spectrum is shown in
Figure~\ref{spectrum}. Two nuclides with presently unknown mass
values~\cite{AW03} are indicated.
%
\subsection{Lifetime Measurement with Single Stored Fragments}
\par
\begin{figure}[t!]
\includegraphics*[width=\textwidth]{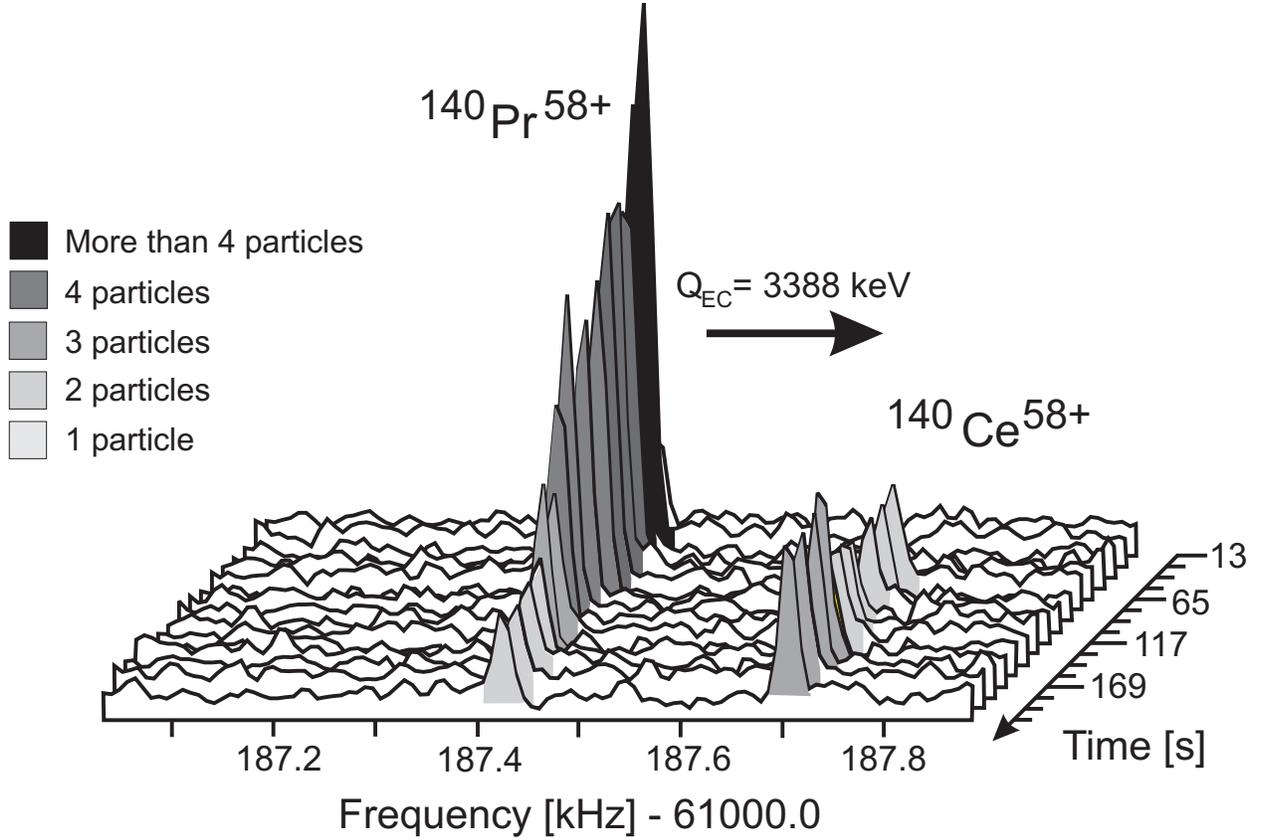}
\caption {A series of subsequent-in-time Schottky frequency
spectra of mother $^{140}$Pr$^{58+}$ and daughter
$^{140}$Ce$^{58+}$ ions. About six mother nuclei were initially
stored. Two out of them decayed via nuclear electron capture into
$^{140}$Ce$^{58+}$. The correlated intensity changes are clearly
seen. Other ions decayed via $\beta^+$ decay or were lost e.g. due
to interaction with the residual gas.} \label{140Pr-decay}
\end{figure}

%
Already in previous experiments we have proven that we are
sensitive and selective to single particles \cite{Li-2005} stored
and cooled in the ESR. SMS is ideally suited to measure decay
properties of bare and few-electron  fragments if the Q-value and
the change in B$\rho$ are not exceeding the storage acceptance of
the ESR \cite{{Ir-PRL},{Li-PLB},{Oh-PRL}}. In case the change of
B$\rho$ values in the decay is too large, the daughter nuclei can
be intercepted by detectors placed outside the storage orbits of
the fragments. In this experiment we aimed at investigating the
decay of a nucleus with a strong electron capture (EC) branch and
a half-life in the order of a few minutes. The selected nucleus
was $^{140}$Pr$^{58+}$ characterized with a Q$_{EC}$ value of 3388
keV. Mother and daughter nuclei are well resolved in the Schottky
spectrum and we specifically aimed our measurements on the study
of single particle decay such that we observed mother and daughter
nuclei discretely changing the area of the corresponding Schottky
frequency peaks. This is really a unique measurement and can be
only performed with our facilities under the described conditions.
An example of measured decays with only a few mother nuclei is
illustrated in Figure~\ref{140Pr-decay} where a series of
subsequent-in-time Schottky frequency spectra are plotted. Goals
of this experimental study are to check the SMS with nuclei of
well-known lifetimes down to a few stored ions and to investigate
the decay statistics under these extreme conditions.
%
\section{Towards pure isomeric beams for the ILIMA project}
\par%
\begin{figure}[b!]
\includegraphics*[width=\textwidth]{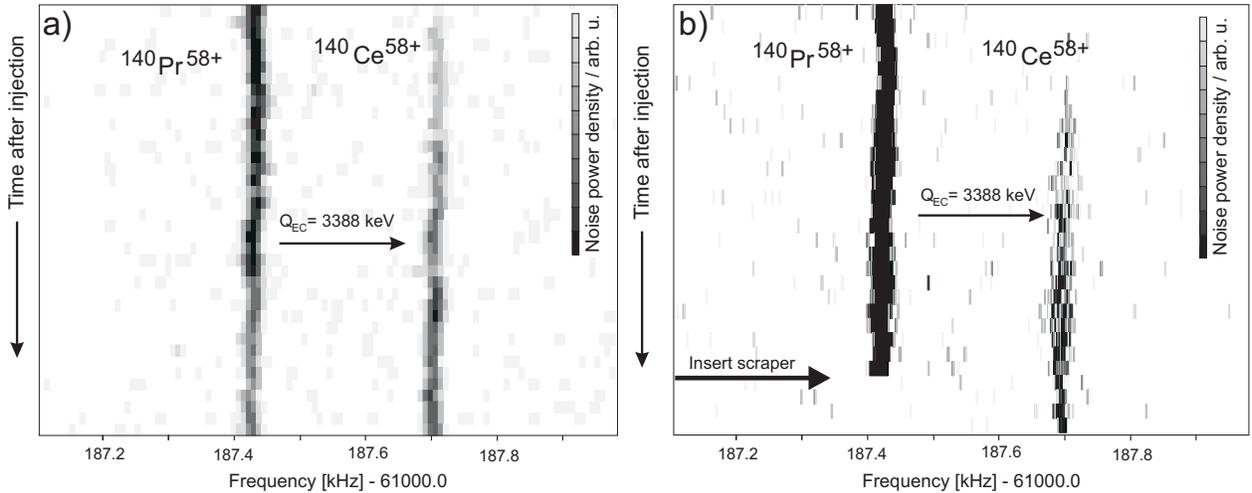}
\caption {Schottky frequency spectra of well resolved mother and
daughter ions characterized by a Q value of about 3 MeV. Note,
that we can inject monoisotopic fragment beams in the ESR as we
have proven many years ago. Left panel: undisturbed mother and
daughter traces in time recorded for about 520~s. Right panel:
170~s after the injection into the ESR, the primary beam of mother
ions was eliminated by mechanical scraping. This is a
demonstration for the feasibility to provide pure isomeric beams
in the storage rings. } \label{ILIMA}
\end{figure}
Although the present experimental program at the SIS-FRS-ESR
facilities has been quite successful and has led to several basic
discoveries, the field of research is expected to be drastically
extended by the next-generation facility FAIR \cite{CDR}. It will
consist of a more powerful driver accelerator, a large acceptance
in-flight separator Super-FRS \cite{Super-FRS}, and a new
storage-cooler ring system specially adapted to the large phase
space and short half-lives of the exotic nuclear beams
\cite{Be-EPAC}. Within the FAIR framework, ILIMA \cite{ILIMA} is
an accepted proposal which will be an extension of the present
successful program at the FRS-ESR. One goal is to provide pure
isomeric beams circulating in the new storage ring system to be
investigated and used in reactions with the internal target or
collision zones with other stored particles as electrons or
antiprotons. An important demonstration of the feasibility in this
direction has been achieved in the present experiment by scraping
off one component of the mother and daughter nuclei recorded with
SMS. The goal was achieved by a precise mechanical scraper at a
dispersive plane in the ESR. This mechanical separation in the
micrometer range brought the success as demonstrated in Figure
\ref{ILIMA}. The technique which has been applied is very similar
to one developed for the measurements of the horizontal beam size
of cooled ion beams with micrometer resolution~\cite{St-EPAC04}.
More sensitive methods to achieve the micrometer separation
\cite{St-EPAC04} can be done by moving the stored beam towards a
fixed position of the scraper.

\section{Conclusion}
\par%
The time-resolved Schottky Mass Spectrometry has again proven its
great potential for precise mass determination of short-lived
nuclides~\cite{Li-2005}. In this work the technique was now
applied to neutron-deficient nuclides below samarium. The covered
mass surface is very close to the astrophysical rp-process path.
Thus our results will contribute to a better understanding of this
nucleosynthesis process.
\par%
Unique results have been achieved in the present experiments with
lifetime measurements of a few mother nuclei stored in the ESR.
\par%
An important step towards the future has been achieved with the
demonstration of a method to provide pure isomeric beams. The
spatial separation of ground states or isomeric states with
excitation energies of at least 3.5~MeV is now a realistic
perspective.


\begin{thebibliography}{99}
\bibitem{Bohr}
A.~Bohr, B.R.~Mottelson, {\it Nuclear Structure}, World Scientific
Publ., Singapore, 1998.
\bibitem{Bender_RMP}
M.~Bender et al., Rev. Mod. Phys. 75 (2003) 121.
\bibitem{Lunney_RMP}
D.~Lunney, J.~Pearson, C.~Thibault, Rev. Mod. Phys. 75 (2003)
1021.
\bibitem{Li-2005}
Yu.A.~Litvinov, H.~Geissel, T.~Radon, et al., Nucl. Phys. A756
(2005) 3.
\bibitem{Li-PRL}
Yu.A.~Litvinov et al., Phys. Rev. Lett. 95 (2005) 042501.
\bibitem{Blasche}
K.~Blasche, B.~Franczak, in Proc.: 3$^{rd}$ Eur. Part. Acc. Conf.,
Berlin, 1992, p. 9.
\bibitem{Ge-NIM24}
H.~Geissel et al., Nucl. Instr. and Meth. B70 (1992) 286.
\bibitem{Franzke}
B.~Franzke, Nucl. Instr. and Meth. B24/25 (1987) 18.
\bibitem{GLOBAL}
C.~Scheidenberger et al., Nucl. Instr. and Meth. B142 (1998) 441.
\bibitem{Raidi}
T.~Radon et al., Nucl. Phys. A677 (2000) 75.
\bibitem{No-NIMA}
F.~Nolden et al., Nucl. Instr. and Meth. A532 (2004) 329.
\bibitem{Ge-RNB6}
H.~Geissel, Yu.A.~Litvinov et al., Nucl. Phys. A746 (2004) 150c.
\bibitem{Borer}
J.~Borer et al., in Proc.: IX$^{th}$ Int. Conf. on High-Energy
Acc., Stanford, 1974, p. 53.
\bibitem{AW03}
A.~Wapstra, G.~Audi, C.~Thibault, Nucl. Phys. A729 (2003) 129.
\bibitem{MOCADI}
N. Iwasa et al., Nucl. Instr. and Meth. B126 (1997) 284.
\bibitem{LISE}
D.~Bazin et al., Nucl. Instr. and Meth. A482 (2002) 307.
\bibitem{GL-2005}
H.~Geissel, Yu.A.~Litvinov, J. Phys. G31 (2005) S1779.
\bibitem{rp}
H.~Schatz et al., private communications.
\bibitem{Ir-PRL}
H.~Irnich, H.~Geissel, F.~Nolden et al., Phys. Rev. Lett. 75
(1996) 4182.
\bibitem{Li-PLB}
Yu.A.~Litvinov et al., Phys. Lett. B573 (2003) 80.
\bibitem{Oh-PRL}
T.~Ohtsubo, L.~Maier, C. Scheidenberger et al., Phys. Rev. Lett.
95 (2005) 052501.
\bibitem{CDR}
An international accelerator facility for beams of ions and
antiprotons, Conceptual Design Report, GSI 2001, {\it
http://www.gsi.de/GSI-Future/CDR/}
\bibitem{Super-FRS}
H.~Geissel et al., Nucl. Instr. and Meth. B204 (2003) 71.
\bibitem{Be-EPAC}
P.~Beller et al., in Proc.: 9$^{th}$ Eur. Part. Acc. Conf.,
Lucerne, Switzerland, 2004, p. 1174, {\it
http://accelconf.web.cern.ch/accelconf/}, and references therein.
\bibitem{ILIMA}
Yu.N.~Novikov, Yu.A.~Litvinov, H.~Weick et al., Study of isomeric
beams, lifetimes and masses, ILIMA, Letter of Intent, GSI, 2004.
\bibitem{St-EPAC04}
M.~Steck et al., in Proc.: 9$^{th}$ Eur. Part. Acc. Conf.,
Lucerne, Switzerland, 2004, p. 1966, {\it
http://accelconf.web.cern.ch/accelconf/}.
%
\end{thebibliography}
\end{document}